\documentclass{article}
\usepackage{amssymb}

\begin{document}
\title{Persistent current and Drude weight in mesoscopic rings}
\author{F. Carvalho Dias,$^{1\ast }$ I. R. Pimentel$^{1}$ and 
M. Henkel$^{2}$ 
  \\ {\small $^{1}$Department of Physics
and CFTC, University of Lisbon} 
   \\ {\small Av. Prof. Gama Pinto 2, 1649-003 LISBOA, Portugal}
\\ {\small $^{2}$Laboratoire de Physique des Mat\'{e}riaux 
(CNRS\ UMR 7556)}
\\ {\small Universit\'{e} Henri Poincar\'{e} Nancy I, B.P. 239}
\\ {\small F-54506 Vand\oe {}uvre l\`{e}s Nancy Cedex, France}}

\date{}
\maketitle

\begin{abstract}
We study the persistent current and the Drude weight of a system of 
spinless fermions, with repulsive interactions and a hopping impurity, 
on a
mesoscopic ring pierced by a magnetic flux, using a Density Matrix
Renormalization Group algorithm for complex fields. Both the Luttinger
Liquid (LL) and the Charge Density Wave (CDW) phases of the system are
considered. Under a Jordan-Wigner transformation, the system is equivalent
to a spin-1/2 XXZ chain with a weakened exchange coupling. We find that 
the
persistent current changes from an algebraic to an exponential decay with
the system size, as the system crosses from the LL to the CDW phase with
increasing interaction $U$. We also find that in the interacting system 
the
persistent current is invariant under the impurity transformation 
$\rho\rightarrow 1/\rho $, for large system sizes, where $\rho $ is the 
defect strength. The persistent current exhibits a decay that is in 
agreement with
the behavior obtained for the Drude weight. We find that in the LL phase 
the
Drude weight decreases algebraically with the number of lattice sites $N$,
due to the interplay of the electron interaction with the impurity, 
while in
the CDW phase it decreases exponentially, defining a localization length
which decreases with increasing interaction and impurity strength. Our
results show that the impurity and the interactions always decrease the
persistent current, and imply that the Drude weight vanishes in the limit 
$N\rightarrow \infty $, in both phases.

\end{abstract}

\newpage

\section{INTRODUCTION}

The experimental discovery of persistent currents in mesoscopic rings
pierced by a magnetic flux,$^{1-3}$ earlier proposed theoretically,$^{4}$
has revealed interesting new effects. The currents measured in metallic 
and semiconducting rings, either in a single ring or an array of many 
rings, generally exhibit an unexpectedly large amplitude, i.e., larger 
by at least
one order of magnitude, than predicted by theoretical studies of electron
models with either disorder or electron-electron interaction treated
perturbatively.$^{5,6}$ It has been suggested that the interactions and
their interplay with disorder are possibly responsible for the large
currents observed, expecting that the effect of the interactions could
counteract the disorder effect. However, no consensus has yet been reached
on the role of the interactions. In order to gain theoretical
insight, it is desirable to perform numerical calculations which allow to
consider both interactions and disorder directly in systems with sizes
varying from small to large. Analytical calculations usually involve
approximations which mainly provide the leading behavior of the properties
for large system sizes. Persistent currents in mesoscopic rings strongly
depend on the system size, since they emerge from the coherence of the
electrons across the entire system. Hence, it is most important to study 
the
size dependence of the current beyond leading order in microscopic models,
for a complete understanding of the experimental results. Exact
diagonalization was used to calculate persistent currents in systems with
very few lattice sites.$^{7,8}$ In this work, we use the Density Matrix
Renormalization Group (DMRG) algorithm,$^{9-11}$ to study a simplified 
model
incorporating interactions and a single impurity, accounting for disorder,
in larger system sizes. We consider a system of interacting spinless
electrons on a one-dimensional ring, with a single impurity, and 
penetrated
by a magnetic field. We study an intermediate range of system sizes, where
analytical results obtained by bosonization techniques for large system
sizes, do not yet fully apply. Without impurity, and at half-filling, the
system undergoes a metal-insulator transition from a Luttinger Liquid 
(LL)$^{12}$ to a Charge Density Wave (CDW)$^{13}$ groundstate. The 
persistent
current of the interacting system with an impurity was studied before with
the DMRG, in the LL phase.$^{14}$ Here we study the persistent current, 
and
also the Drude weight characterizing the conducting properties of the
system, in both the LL and the CDW phase, investigating the interplay
between the impurity and the interactions in the two phases. In mesoscopic
systems the separation between metallic and insulating behavior is not
always obvious, since\ the localization length can be of the order or
significantly larger than the system size. Hence, a finite Drude weight 
and
a current can be observed in the CDW phase of a mesoscopic system. It is
therefore of great interest to characterize the persistent current and the
Drude weight in both the LL and the CDW phases of mesoscopic systems.
Although the simple model that we consider is not the most appropriate to
describe the experimental situation, we hope to obtain useful information
for the understanding of the more realistic systems. Under a Jordan-Wigner
transformation,$^{15}$ the system considered is equivalent to a spin-1/2 
XXZ
chain with a weakened exchange coupling. Hence, our results also provide
insight into the spin transport in this type of systems.

\section{THE MODEL}

The Hamiltonian describing a system of spinless fermions on a ring pierced
by a magnetic flux, with repulsive interactions and a single hopping
impurity, or defect, is given by, 
\begin{equation}
H=H_{t}+H_{U},  \label{1}
\end{equation}
where 
\begin{equation}
H_{t}=-t\sum_{j=1}^{N}\left( e^{i\phi /N}c_{j}^{\dagger }c_{j+1}+e^{-i\phi
/N}c_{j+1}^{\dagger }c_{j}\right) +(1-\rho )t\left( e^{i\phi
/N}c_{N}^{\dagger }c_{1}+e^{-i\phi /N}c_{1}^{\dagger }c_{N}
\right)  \label{2}
\end{equation}
is the hopping term, $\phi =2\pi \Phi /\Phi _{o}$ contains the magnetic 
flux 
$\Phi $ in units of the flux quantum $\Phi _{o}=hc/e$, $\rho $ measures 
the
strength of the defect with values between $0$ and $1$, ($\rho =1$
corresponding to the defectless case), and 
\begin{equation}
H_{U}=U\sum_{j=1}^{N}n_{j}n_{j+1}  \label{3}
\end{equation}
is the interaction term, with $U>0$ representing the nearest neighbor
Coulomb repulsion, and $n_{j}=c_{j}^{\dagger }c_{j}$, where 
$c_{j}^{\dagger}$ and $c_{j}$ are the spinless fermion operators 
acting on the site $j$ of
the ring. We consider a system of $N$ sites, with $N$ even, and at
half-filling, when $M=N/2$ particles are present. The lattice constant is
set to one and periodic boundary conditions, $c_{N+1}=c_{1}$, are used.

Via the gauge transformation $c_{j}\rightarrow e^{-i\phi j/N}c_{j}$, the
flux can be removed from the Hamiltonian, but in the impurity term where 
the
flux is trapped, and the quantum phase $\phi $ is encoded in a twisted
boundary condition $c_{N+1}=e^{-i\phi }c_{1}$. It is then clear that the
energy is periodic in $\phi $ with period $2\pi $, i.e., it is periodic in
the flux $\Phi $ threading the ring with period $\Phi _{0}$.$^{16}$

After a Jordan-Wigner transformation, Eqs. (2) and (3) can be rewritten,
respectively, as 
\begin{equation}
H_{J}=-\frac{J}{2}\sum_{j=1}^{N}\left(
S_{j}^{+}S_{j+1}^{-}+S_{j}^{-}S_{j+1}^{+}\right) +(1-\rho )
\frac{J}{2}\left(e^{i\phi }S_{1}^{+}S_{N}^{-}+e^{-i\phi }S_{1}^{-}
S_{N}^{+}\right)  \label{4}
\end{equation}
and 
\begin{equation}
H_{\Delta }=\Delta \sum_{j=1}^{N}S_{j}^{z}S_{j+1}^{z}  \label{5}
\end{equation}%
with $t=J/2$ and $U=\Delta $, and the boundary conditions 
$S_{N+1}^{+}=(-1)^{M+1}e^{i\phi }S_{1}^{+}$ and $S_{N+1}^{z}=S_{1}^{z}$.
Hence, the model (1) of spinless fermions is equivalent to a spin-1/2 XXZ
chain with a weakened exchange coupling, and twisted boundary conditions 
in the transverse direction. The half-filled case corresponds to total 
spin projection $S^{z}=0$.

The persistent current generated on a ring pierced by a magnetic flux, at
temperature $T=0$, can be obtained from the ground state energy 
$E_{o}(\phi)$, by taking the derivative with respect to $\phi$, 
\begin{equation}
I(\phi )=-\frac{\partial E_{o}(\phi )}{\partial \phi }.  \label{6}
\end{equation}

\noindent For the spinless fermion system, Eqs. (2) and (3), $I(\phi )$
corresponds to the ground state value of the charge current operator 
$\hat{I}_{c}=it\sum_{j=1}^{N}\left( c_{j}^{\dagger }c_{j+1}-
c_{j+1}^{\dagger}c_{j}\right)$, while for the XXZ chain, Eqs. (4) 
and (5), it corresponds to the ground state value of the spin current 
operator $\hat{I}_{s}=i\frac{J}{2}\sum_{j=1}^{N}\left( S_{j}^{+}
S_{j+1}^{-}-S_{j+1}^{+}S_{j}^{-}\right)$.
As a consequence of the periodicity of the energy, the current is also
periodic in $\phi $, with period $2\pi $. Hence, it can be expressed as a
Fourier series,

\begin{equation}
I(\phi )=\sum_{k=1}^{\infty }I_{k}\sin (k\phi ).  \label{7}
\end{equation}

\noindent and the behavior of the current can be analyzed in terms of the
coefficients $I_{k}$.$^{14}$

In the noninteracting case ($U=0$), it has been found that for large 
system sizes, the current is invariant under the defect transformation 
$\rho\rightarrow 1/\rho $,$^{17,18}$ i.e.,

\begin{equation}
I(\rho )=I(1/\rho ).  \label{8}
\end{equation}

\noindent We shall investigate the existence of this kind of invariance in
the interacting case ($U>0$), both in the LL and the CDW phases.

The Drude weight was proposed by Kohn as a relevant quantity to 
distinguish
between a metal and an insulator.$^{19}$ It is defined as

\begin{equation}
D=N\frac{\partial ^{2}E_{o}}{\partial \phi ^{2}}\left\vert _{\phi =\phi
_{m}}\right. ,  \label{9}
\end{equation}
where $\phi _{m}$ is the location of the minimum of $E_{o}(\phi )$, which
depends on the parity of the number of electrons, i.e., $\phi _{m}=0$ or 
$\pi $ for, respectively, an odd or an even number of electrons. For the
spinless fermion system $D$ represents the charge-stiffness and 
measures the
inverse of the effective mass of the charge carriers.$^{20}$ In a metallic
conductor $D$ tends to a finite value whereas in an insulator $D$ vanishes
with the system size $N$, when $N\rightarrow \infty $. In the insulating
state the Drude weight decays as $D\sim \exp (-N/\xi )$, where $\xi $
measures the localization length. For the XXZ chain the Drude weight
represents the spin-stiffness.

In a model of free fermions ($U=0$) with no impurity ($\rho =1$), it is
straightforward to see that the leading behavior of the persistent 
current 
$I(\phi )$ in the system size $N$, has a saw-tooth like shape with slope 
$-v_{F}/(\pi N)$, where $v_{F}$ is the Fermi velocity. Thus, the 
amplitude of
the current scales with $1/N$, vanishing in the limit 
$N\rightarrow \infty$.
 The discontinuity in $I(\phi )$, that results from a degeneracy of energy
levels associated to the translation symmetry,$^{8}$ appears at 
$\phi =0$ or 
$\pi $ for, respectively, an even or an odd number of electrons. In the
presence of an impurity ($\rho \neq 1$), bosonization$^{21}$ and conformal
field theory$^{17}$ calculations predict that the shape of the current 
$I(\phi )$ is rounded off, and its amplitude decreases with increasing
strength of the scatterer potential, still vanishing with the 
system size as $1/N$. The impurity lifts the degeneracy of the energy 
levels and the current then varies continuously.$^{8}$

The model with interactions ($U\neq 0$) and without defect ($\rho =1$), is
solvable by the Bethe ansatz for periodic boundary conditions 
($\phi =0$)$^{22,23}$ and also for twisted boundary conditions 
($\phi \neq 0$).$^{24-26}$
At half-filling, the system exhibits a metal-insulator transition, which
occurs at $U/t=2$. For $U/t<2$, the system is in a gapless LL phase, while
for $U/t>2$ it is in a gapped CDW state.$^{27}$ The LL phase is
characterized by a power-law decay of the correlations. Bosonization
predicts that in an homogeneous LL, the leading behavior of the persistent
current in the system size $N$, has a saw tooth like shape with slope 
$-v_{J}/(\pi N)$, where $v_{J}$ is the velocity of current 
excitations.$^{28}$ Since translation invariance is preserved in the 
presence of interactions,
the discontinuity in the current still exists for finite $U$.$^{8}$ 
A Bethe
ansatz calculation shows that the Drude weight of an homogeneous LL in the
thermodynamic limit, has a finite value, which decreases with increasing
strength of the interaction $U/t$.$^{20}$ The LL state is strongly 
affected
by the presence of an impurity,$^{29-32}$ and bosonization yields that the
current then vanishes as $I\sim N^{-1-\alpha _{B}}$, with 
$\alpha _{B}>0$.$^{21}$ The study of the LL phase with $\rho \neq 1$, 
performed with the DMRG,$^{14}$ has in fact found this kind of 
behavior. The CDW phase is
characterized by a localization length $\xi $, which is associated to the
energy gap. From the work of Baxter,$^{33}$ the Drude weight in the gapped
phase, is expected to behave as $D\sim \exp (-N/\xi )$, vanishing for an
infinite system size. This behavior implies that although the system is
insulating in the infinite system size limit, for a finite system, 
provided 
$N/\xi $ is not too large, $D$ is still finite and a current can be 
observed.
The localization length can then be extracted from the size dependence of
the Drude weight.

\section{NUMERICAL METHOD}

We use the DMRG to numerically calculate the groundstate energy of the
spinless fermion system as a function of the magnetic flux, 
$E_{o}(\phi )$,
for fixed interaction $U$ and impurity strength $\rho $, in rings up to 
$N=82 $ sites, keeping up to $250$ density matrix eigenstates per 
block.$^{34} $ The DMRG is applied to the Hamiltonian (1) after performing
the gauge transformation, which removes the flux into a twisted 
boundary condition.$^{11}$ The states of the system are characterized by the 
quantum numbers
associated to the eigenvalues of the local occupation number $n_{j}$ and 
the
total number of particles $M=\sum_{j=1}^{N}n_{j}$ operators, which commute
with the Hamiltonian (1).$^{35}$ For each set of $N$, $\rho $ and $U$, we
obtained the groundstate energy $E_{o}$ for $50$ values of $\phi $ in the
periodicity interval $-\pi <\phi \leq \pi $, and using Chebyshev
interpolation$^{36}$ we determined the corresponding current (6) and Drude
weight (9), by numerical differentiation. We developed a DMRG algorithm 
for
complex Hamiltonian matrices, which allowed to calculate the detailed form
of the persistent current $I$ as a function of $\phi $,$^{14}$ and to 
obtain
the Drude weight. In a previous approach the DMRG was used to calculate 
the
so called phase sensitivity $\Delta E_{0}$, which is the difference of the
groundstate energy at flux $\phi =0$ and $\pi $, and can be considered a
crude measure of the persistent current.$^{37-39}$ Although the 
calculation
of $\Delta E_{0}$ requires considerably less computational effort than the
calculation of $E_{0}(\phi )$, because then the Hamiltonian matrix is 
real,
the phase sensitivity does not provide information on the shape of the
current and the value of the Drude weight.

\section {RESULTS}

We now present the results obtained for the persistent current and the 
Drude
weight, where we take $t=1$ and the interaction $U$ is in units of $t$.
Figs. 1 and 2, exhibit $NI(\phi )$ plotted versus $\phi $, for 
respectively, 
$U=0.80$ and $U=3.00$, which correspond, respectively, to the LL and the 
CDW
phase, considering different impurity strengths $\rho $, on a fixed system
size $N=26$. We can see that the effect of the impurities, in both phases,
is to reduce the intensity of the current, and to round off the shape of 
$I(\phi )$. The amplitude of the current decreases rapidly with increasing
values of $|\rho -1|$, and also with increasing strength of the 
interaction 
$U$. Fig. 3 shows that the invariance of the current with respect to the
defect, described in Eq. (8), found for large system sizes in the
noninteracting case, is also observed for the interacting case, both in 
the
LL and in the CDW phases, the system size $N$ required to reach the
invariance being larger for larger interaction $U$. Fig. 4 displays the
Drude weights associated to the systems with different interactions $U$ 
and
impurities $\rho $, fixed $N=26$, of the currents presented in Figs. 1 and
2. As one would expect the Drude weight decreases with increasing 
$|\rho -1|$, and also increasing $U$. Figs. 5 and 6, present $NI(\phi )$ 
plotted for
several system sizes $N$, respectively, for $U=0.80$, in the LL phase, 
and 
$U=3.00$, in the CDW phase, with $\rho $ fixed at $0.50$. We observe 
that the
current vanishes faster than $1/N$ in both phases, exhibiting a different
behavior in each phase, $I(\phi )$ vanishing much faster with $N$ 
in the CDW
than in the LL phase. In order to analyze the behavior of the current in
more detail we have numerically evaluated the coefficients $I_{k}$ (for 
$k=1,2$) of the Fourier expansion (7). The first ($k=1$) and the second 
($k=2$) Fourier coefficients of the current, for $U=0.80$ and 
$U=3.00$, are shown
in Fig. 7. One can clearly see that the coefficients $I_{1}$ and $I_{2}$
behave similar to each other in both phases. However, their behavior in 
the
LL and CDW phase is distinct. In the LL phase, the Fourier coefficients 
show
a power-law decay with $N$, with the second order coefficient decaying
faster, i.e., with a larger exponent, than the first one. In the CDW 
phase,
the Fourier coefficients show a dominant exponential decay with $N$, with
the second order coefficient also decaying faster, i.e., with a smaller
localization length, than the first one. We observe that for longer rings,
stronger interactions and also stronger impurities, the current is
increasingly more precisely described by its first Fourier component. In 
the
LL phase, this in fact corresponds to the asymptotic behavior predicted by
bosonization in the large $N$\ limit, that is 
$I\sim N^{-1-\alpha _{B}}\sin\phi $. However, for the system sizes 
considered here this asymptotic regime
is not reached and the current displays a more complex behavior. 
The current
is composed of a few Fourier components with decreasing weight. Fig. 8
presents the first Fourier coefficient of the current for different values
of the interaction, $U=0.80$, $2.50$, $3.00$, fixed $\rho =0.50$, from 
which
we extract the dependence of $I_{1}$ on $N$, in the intermediate range of
sizes considered. We observe that for $U=0.80$, in the LL phase, the first
coefficient of the current varies as $I_{1}\sim $ $N^{-1-\alpha _{1}}$, 
with 
$\alpha _{1}\simeq 0.06$, while for $U=2.50$ and $U=3.00$, in the CDW 
phase, it varies as $I_{1}\sim N^{-1-\delta _{1}}\exp (-N/\xi _{1})$,
respectively, with $\xi _{1}\simeq 259$, $\delta _{1}=0.11$, and $\xi
_{1}\simeq 68$, $\delta _{1}=0.10$. The exponent $\alpha _{1}$ is given by
the slope of the straight line in Fig. 8.a, and the length $\xi _{1}$ and
the exponent $\delta _{1}$ were carefully adjusted in order to obtain the
best collapse of the data in Fig.8.b, on a plot of $\ln (N^{1+\delta
_{1}}I_{1})$ vs $N/\xi _{1}$. The Drude weights characterizing the systems
with different interactions $U$, fixed $\rho =0.50$, are presented in Fig.
9. Fig. 10 clearly shows that the results obtained for the Drude weight
confirm the conducting behavior shown by the first coefficient of the
currents in Fig.8. We observe that, for $U=0.80$ the Drude weight varies with
the system size as $D\sim N^{-\alpha }$, with $\alpha \simeq 0.04$, while
for $U=2.50$ and $U=3.00$ it varies as $D\sim N^{-\delta }\exp (-N/\xi )$,
respectively, with $\xi \simeq 307$, $\delta \simeq 0.08$ and 
$\xi \simeq 68$, $\delta \simeq 0.06$. The exponent $\alpha $ is given by 
the slope of the
straight line in Fig. 10.a, and the localization length $\xi $ and the
exponent $\delta $ were carefully adjusted in order to obtain the best
collapse of the data in Fig.10.b, on a plot of $\ln (N^{\delta }D)$ vs 
$N/\xi $. The exponents and localization lengths characterizing the Drude
weight are a little different from those characterizing the first Fourier
component of the current, as one would expect, since the Drude weight
contains the contribution from the various Fourier components. One sees 
that
the localization length $\xi $ and the exponent $\delta $ decrease with
increasing strength of the interaction $U$. Also, concerning the impurity
influence, Fig. 4 implies that the exponent $\alpha $ in the LL phase
increases, and the localization length $\xi $ in the CDW phase decreases,
with increasing $\left\vert \rho -1\right\vert $. As mentioned, in the 
large 
$N$ limit the current is expected to behave as its first 
Fourier component,
which in the LL phase implies that the exponent $\alpha _{1}$ should be
identified with $\alpha _{B}=1/K-1$ as calculated from 
bosonization,$^{21}$
where $K$ is the LL parameter, calculable from the Bethe 
ansatz.$^{40}$ For 
$U=0.80$ this leads to $\alpha _{B}\simeq 0.27$, which is much larger than
our value of $\alpha _{1}$. We should note that the size dependence found
for the first Fourier component of the current, and the Drude weight,
characterizes the behavior of an intermediate and limited range of system
sizes. If one would consider a larger range of systems, in the LL 
phase, one
would most probably see the data for the larger $N$ bending down, crossing
to an asymptotic power-law behaviour with the exponent approaching 
$\alpha_{B}$. This was observed in Ref. $14$, where the behavior of 
the first few
Fourier components of the current in the LL phase was discussed in detail,
with data taken for larger values of $N$ and stronger interaction and
impurity strengths. Also, in the CDW phase we consider systems in an
intermediate regime where the localization length is larger or near the
system size.$^{41}$ For larger systems, the power factors that occur in 
the
first Fourier component of the current and the Drude weight may 
decline,$^{42}$ possibly leaving a pure exponential behavior in 
the asymptotic regime.

From the results obtained, we observe that the system with $U=0.80$ and 
$\rho =0.50$, is characterized by an exponent $\alpha >0$, which is 
generated
by the interplay of the electron interaction with the impurity, and $D$
exhibits then a power-law decay with $N$, which implies vanishing in the
limit $N\rightarrow \infty $. On the other hand, the systems with $U=2.50$
and $U=3.00$, fixed $\rho =0.50$, are characterized by a localization 
length 
$\xi $, which decreases with increasing interaction and impurity strength,
and $D$ exhibits now an exponential decay with $N$, also vanishing as 
$N\rightarrow \infty $. Hence, we find that both in the LL and the CDW
phases, with an impurity in the system, the effect of the interaction is 
to
decrease the current and the Drude weight. As referred before, our results
also provide insight into the spin transport in a spin-1/2 XXZ chain 
with a
weak link, and a similar behavior to the one above is implied for the spin
current and stiffness. So, in the gapless XY phase, the spin stiffness
decays with a power-law, vanishing in the limit $N\rightarrow \infty $,
while in the gapped Ising phase, it decays exponentially, also 
vanishing as 
$N\rightarrow \infty $. Comparing our results for the persistent current 
in
the LL phase, with those obtained in Ref. $14$, we have similarly found 
that
 the current vanishes faster than $1/N$. One
observes that the model parameters strongly influence when the last
asymptotic regime described by bosonization is reached. A calculation of 
the
finite-size corrections to the spin stiffness in a pure spin-1/2 XXZ 
chain,$^{42-44}$ has revealed a size dependence in the gapped phase that 
has a
similar form to the one found here. The result that the Drude weight in a
ring in the gapless phase with an impurity drops to zero, is in agreement
with a previous result obtained for a spin chain,$^{38}$ and with
renormalization group arguments, which state that the impurity term is
relevant leading to a transmission cut.$^{31,32}$ The renormalization 
group
studies find that either a weak barrier or a weak link lead to an 
insulating
state for repulsive interactions, while in the noninteracting case those 
are
marginal perturbations. In turn, our work shows that there is an 
invariance
of the current under the defect transformation 
$\rho \rightarrow 1/\rho $ in
the interacting system, as for the noninteracting system, and that implies
that a strong link will also reduce the current and the Drude weight. The
observation that with an impurity in the system, the interaction always
leads to an additional decrease of the current and the Drude weight is in
agreement with previous results by other authors,$^{7,8}$ and can be
understood as it is more difficult to move correlated electrons in a
scattering potential than independent electrons.

\section{SUMMARY}

We have studied the behavior of the persistent current and the Drude 
weight
on a mesoscopic ring pierced by a magnetic flux. We considered a model of
spinless fermions with repulsive interactions and a hopping impurity, 
which
is also equivalent to a \ spin-1/2 XXZ chain with a weakened exchange
coupling. Using a powerful numerical method, the DMRG with complex fields,
we have calculated the detailed form of the current as a function of the
magnetic flux, which enabled us to investigate the corrections to the 
large
system-size limit, and also allowed to obtain the Drude weight. We show 
that
the system at half-filling, changes from an algebraic to an exponential behavior as the
interaction increases, corresponding to a change from a LL to a CDW phase.
We find that the analytical predictions of bosonization for the LL phase,
are not yet fully observed in the intermediate range of system sizes
considered. In addition we observe that the invariance of the current 
under
the defect transformation $\rho \rightarrow 1/\rho $, seen in the
noninteracting system, is also verified in the interacting system, in both
phases. Hence, an isolated strong link is not only useless for increasing
the persistent current (as might have been expected), but it rather 
destroys
coherence and reduces the current. The behavior determined for the 
current
is consistent with the behaviour determined for the Drude weight, the LL
phase being characterized by an exponent $\alpha >0$, which results from 
the
interplay of the interactions with the impurity, while the CDW phase is
characterized by a localization length $\xi $, which decreases with
increasing interaction and impurity strength. We find that, both in the LL
and the CDW phase, with a defect in the system the interactions always
suppress the current, and the Drude weight drops to zero in the limit 
$N\rightarrow \infty $. Away from half-filling there is no metal-insulator 
transition in the pure case, and the system is always metallic. Hence, one 
does not expect to observe then a change in the current and the Drude weight 
from an algebraic to an exponential decay. Nevertheless, one still 
expects to observe that in the system with an impurity, the current and 
the Drude weight decrease with increasing impurity and interaction 
strengthes. Therefore, within the model considered, the 
interactions
cannot explain the results observed in the experiments.

\bigskip

\noindent $^{*}$Electronic address: fdias@cii.fc.ul.pt

\bigskip

\noindent {\LARGE References:}

\medskip

\noindent $^{1}$L.P. L\'{e}vy, G.\ Dolan, J. Dunsmuir, and H. Bouchiat,
Phys. Rev. Lett. \textbf{64}, 2074 (1990).

\noindent $^{2}$V. Chandrasekhar, R.A. Webb, M.J. Brady, M.B. Ketchen, 
W.J. Gallagher, and A. Kleinsasser, Phys. Rev. Lett. \textbf{67}, 
3578 (1991).

\noindent $^{3}$D. Mailly, C. Chapelier, and A. Benoit, Phys. Rev. Lett. 
\textbf{70}, 2020 (1993).

\noindent $^{4}$M. B\"{u}ttiker, Y. Imry, and R. Landauer, Phys. Lett. 
\textbf{96A}, 365 (1983).

\noindent $^{5}$Y. Imry, 
\textit{Introduction to Mesoscopic Physics}, Oxford 
University Press (1997).

\noindent $^{6}$U. Eckern and P. Schwab, Adv. Phys. \textbf{44}, 
387 (1985); J. of Low Temp. Physics \textbf{126}, 1291 (2002).

\noindent $^{7}$M. Abraham and R. Berkovits, Phys. Rev. Lett. \textbf{70},
1509 (1983).

\noindent $^{8}$G. Bouzerar, D. Poilblanc, and G. Montambaux, 
Phys. Rev. B \textbf{49}, 8258 (1994).

\noindent $^{9}$S.R. White, Phys. Rev. Lett. \textbf{69}, 2863 (1992).

\noindent $^{10}$I. Peschel, X. Wang, M. Kaulke, and K. Hallberg (Eds.), 
\textit{Density-Matrix Renormalization - A New Numerical Method in 
Physics}, Springer (1999).

\noindent $^{11}$U. Schollw\"{o}ck, Rev. Mod. Phys. \textbf{77}, 
259 (2005).

\noindent $^{12}$J. Voit, Rep. Prog. Phys. \textbf{58}, 977 (1995).

\noindent $^{13}$See, e.g., G. Gr\"{u}ner, \textit{Density Waves in 
Solids}, Addison-Wesley Publishing Company (1994).

\noindent $^{14}$V. Meden and U. Schollw\"{o}ck, Phys. Rev. B \textbf{67},
035106 (2003).

\noindent $^{15}$See, e.g., G. D. Mahan, \textit{Many-Particle Physics},
Plenum Press (1990).

\noindent $^{16}$N. Byers and C.N.Yang, Phys. Rev. Lett. \textbf{7}, 46
(1961).

\noindent $^{17}$M. Henkel and D. Karevski, Eur. Phys. J. B \textbf{5}, 
787 (1998).

\noindent $^{18}$M. Henkel, \textit{Conformal invariance and Critical
Phenomena}, Springer (1999).

\noindent $^{19}$W. Kohn, Phys. Rev. \textbf{133}, A171 (1964).

\noindent $^{20}$B.S. Shastry and B. Sutherland, Phys. Rev. Lett. 
\textbf{65}, 243 (1990).

\noindent $^{21}$A.O. Gogolin and N.V. Prokof'ev, Phys. Rev. B 
\textbf{50}, 4921 (1994).

\noindent $^{22}$C.N. Yang and C.P. Yang, Phys. Rev. \textbf{150}, 321
(1966).

\noindent $^{23}$J. Des Cloizeaux and Michel Gaudin, J. Math. Phys. 
\textbf{7}, 1384 (1966).

\noindent $^{24}$C.J. Hamer, G.R.W. Quispel, and M.T. Batchelor, 
J. Phys. A \textbf{20}, 5677 (1987).

\noindent $^{25}$F.C. Alcaraz, M.N. Barber, and M.T. Batchelor, Phys. Rev.
Lett. \textbf{58}, 771 (1987); Ann. Phys. \textbf{182}, 280 (1988).

\noindent $^{26}$B. Sutherland and B.S. Shastry, Phys. Rev. Lett. 
\textbf{65}, 1833 (1990).

\noindent $^{27}$In the XXZ\ chain, a transition occurrs at $\Delta /J=1$,
from an XY model, for $\Delta /J<1$, to an Ising model, for 
$\Delta /J>1$, $\Delta /J=1$ corresponding to the Heisenberg model. 
A gap opens up in the spin-excitation spectrum, that is equivalent to 
the gap in the charge-excitation spectrum of the spinless fermion system.

\noindent $^{28}$F.D.M. Haldane, J.Phys. C \textbf{14}, 2585 (1981).

\noindent $^{29}$W. Apel and T.M. Rice, Phys. Rev. B \textbf{26}, R7063
(1982).

\noindent $^{30}$T. Giamarchi and H.J. Schulz, Phys. Rev. B \textbf{37}, 
325 (1988).

\noindent $^{31}$C.L.Kane and M.P.A. Fisher, Phys. Rev. B \textbf{46}, 
15233 (1992).

\noindent $^{32}$S.Eggert and I. Affleck, Phys. Rev. B \textbf{46}, 10866
(1992).

\noindent $^{33}$R.J. Baxter, \textit{Exactly Solved Models in Statistical
Mechanics, }Academic Press, (1982).

\noindent $^{34}$The number $m$ of eigenstates required to achieve
convergence, of 1 part in $10^{5}$, for the ground state energy $E_{0}$,
increases with the system size, e.g., $m=128$ for $N=30$ and $m=250$ for 
$N=66$.

\noindent $^{35}$For the spin XXZ chain the quantum numbers characterizing
the states correspond to the local spin projection $S_{j}^{z}$ and the 
total spin projection $S^{z}=\sum_{j=1}^{N}S_{j}^{z}$.

\noindent $^{36}$W.H. Press, S.A. Teukolsky, W.T. Vetterling, and B.P.
Flannery, \textit{Numerical Recipies in Fortran 77 - The Art of Scientific
Computing (Second Edition)}, Cambridge University Press (2001).

\noindent $^{37}$P. Schmitteckert, T. Schulze, C. Schuster, P. Schwab, and
U. Eckern, Phys. Rev. Lett. \textbf{80}, 560 (1998).

\noindent $^{38}$D. Weinmann, P. Schmitteckert, R.A. Jalabert, and J.L.
Pichard, Eur. Phys. J. B \textbf{19}, 139 (2001).

\noindent $^{39}$T.M.R. Byrnes, R.J. Bursill, H.-P. Eckle, C.J. Hamer and
A.W. Sandvik, Phys. Rev. B \textbf{66}, 195313 (2002).

\noindent $^{40}$F.D.M. Haldane, Phys. Rev. Lett. \textbf{45}, 1358 
(1980).

\noindent $^{41}$N. Laflorencie and H. Rieger, Eur. Phys. J. B 
\textbf{40}, 201 (2004).

\noindent $^{42}$S-J. Gu, V.M. Pereira and N.M.R. Peres, Phys. Rev. B 66,
235108-1, (2002).

\noindent $^{43}$N. Laflorencie, S. Caponni, and E.S. S\o rensen, 
Eur. Phys. J. B \textbf{24}, 77 (2001).

\noindent $^{44}$F. Heidrich-Meisner, A. Honecker, D.C. Cabra, and W.
Brenig, Phys. Rev. B \textbf{68}, 134436 (2003).

\newpage

\noindent {\large Figure Captions:}

\medskip

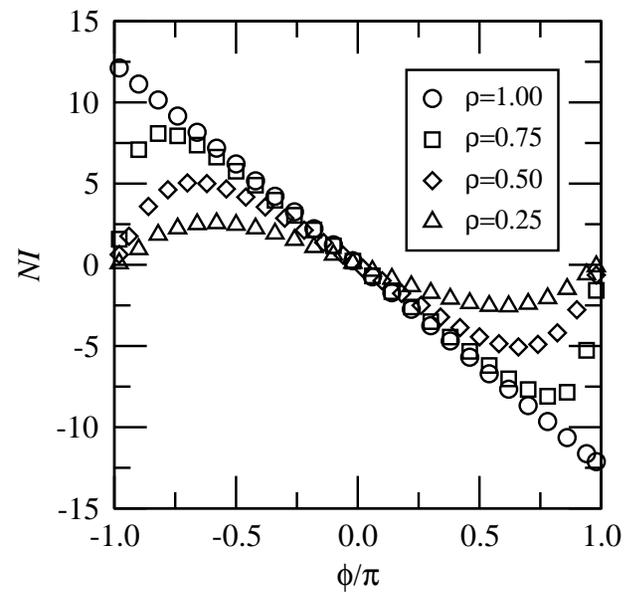
\noindent FIG. 1. Persistent current $NI(\phi )$ vs $\phi $, at fixed 
$N=26$, for $U=0.80$ and different $\rho $.

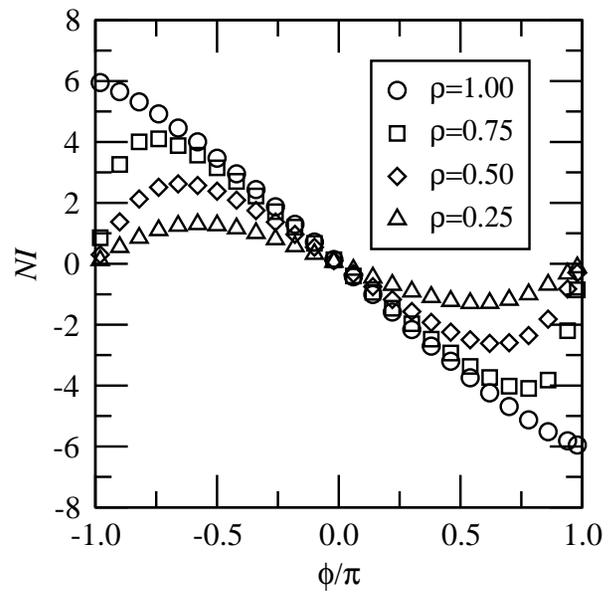
\noindent FIG. 2. The same as in Fig. 1, but for $U=3.00$.

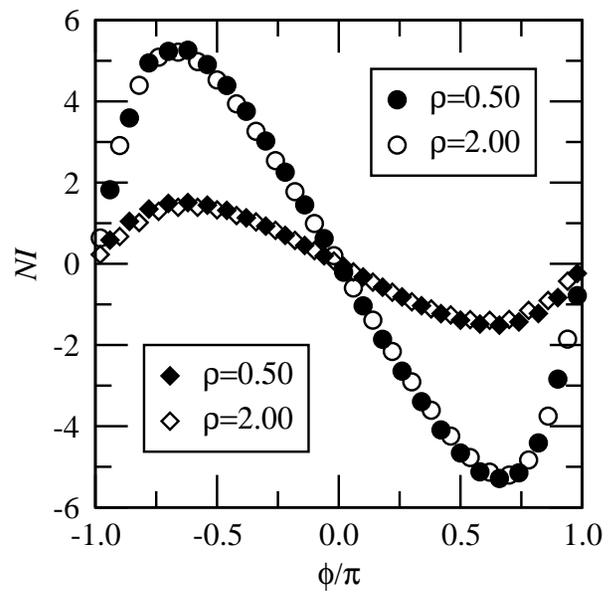
\noindent FIG. 3. Symmetry of the current $NI(\phi )$ in $\rho $ (empty
symbols) vs $1/\rho $ (filled symbols), for $U=0.80$ (circles) at $N=18$,
and $U=3.00$ (diamonds) at $N=58$.

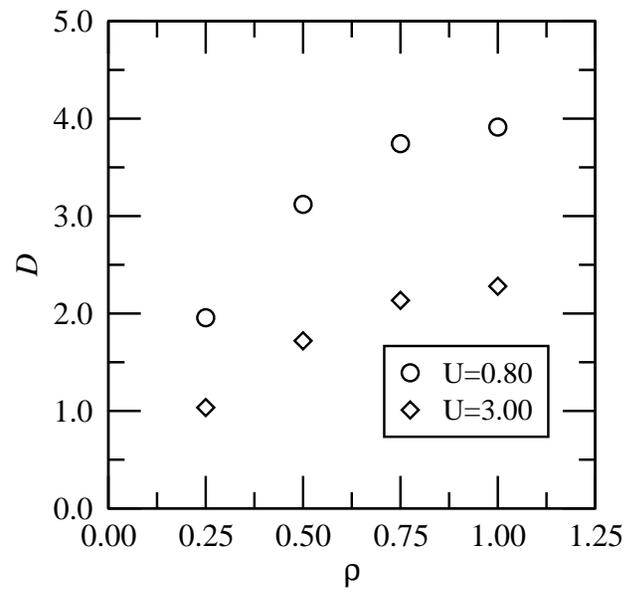
\noindent FIG. 4. The Drude weigth $D$ vs $\rho $, at fixed $N=26$, for 
$U=0.80$ and $U=3.00$.

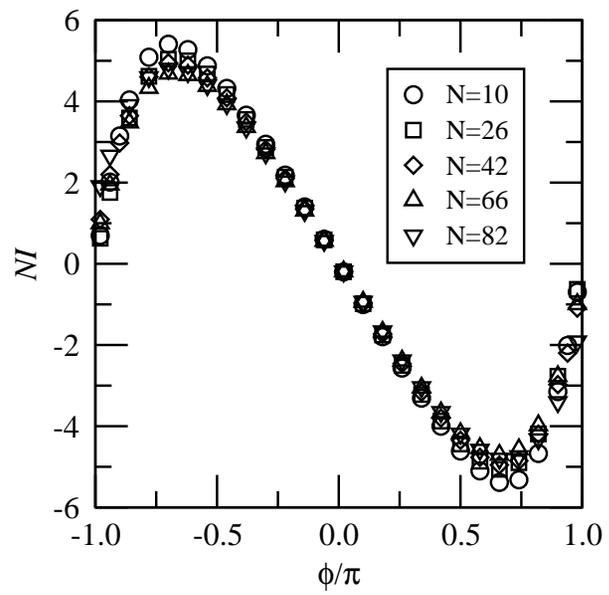
\noindent FIG. 5. Persistent current $NI(\phi )$ vs $\phi $, for 
$U=0.80$, $\rho =0.50$ and increasing $N$.

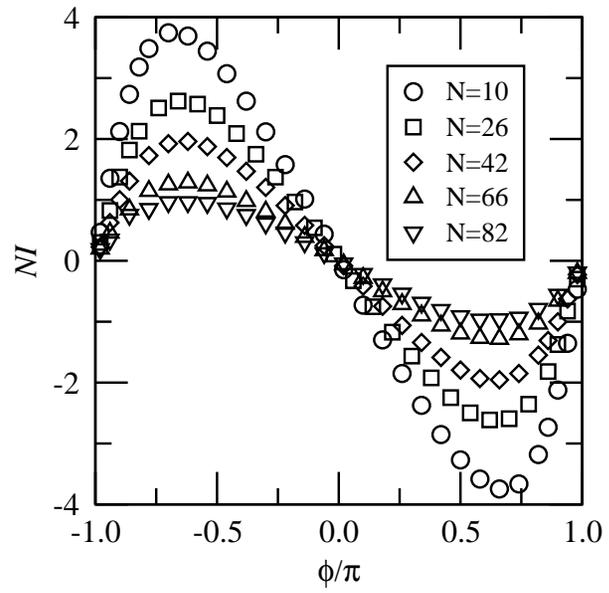
\noindent FIG. 6. The same as in Fig. 5, but for $U=3.00$.

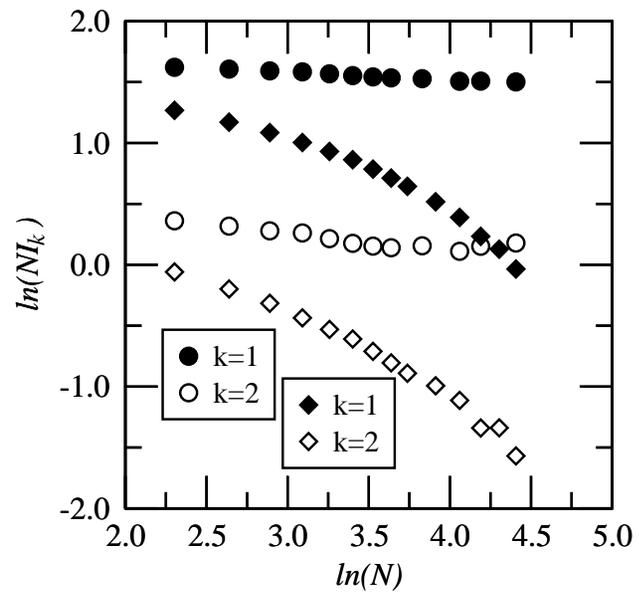
\noindent FIG. 7. Fourier coefficients of the current. $\ln (NI_{k})$ vs 
$\ln (N)$, for $U=0.80$ (circles) and $U=3.00$ (diamonds), at 
$\rho =0.50$, for $k=1$ (filled symbols) and $k=2$ (empty symbols).

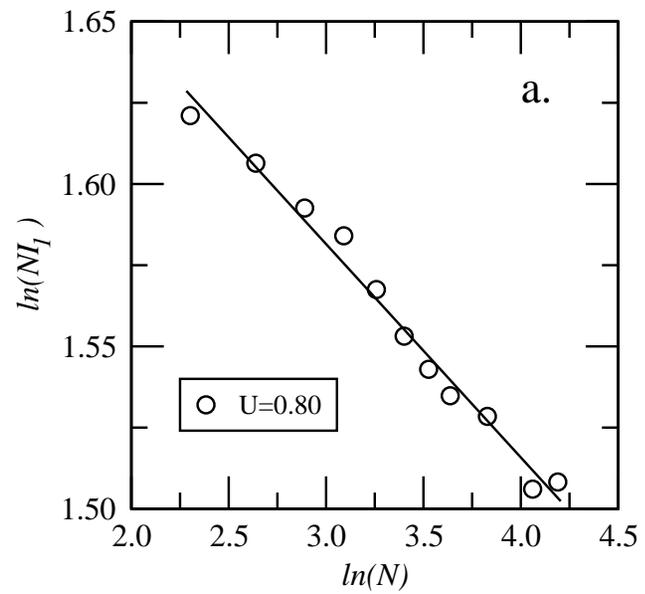
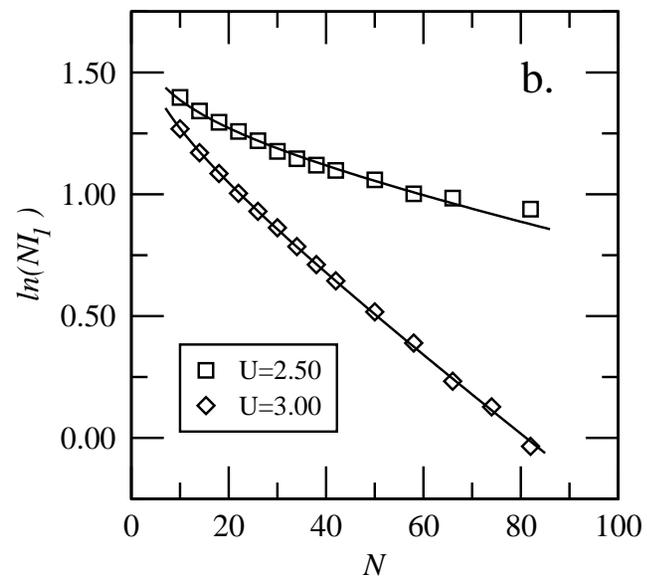
\noindent FIG. 8. $NI_{1}$ for different $U$ and $\rho =0.50$. (a) 
$\ln(NI_{1})$ vs $\ln (N)$, for $U=0.80$, the line represents 
$I_{1}\sim N^{-1-\alpha _{1}}$. (b) $\ln (NI_{1})$ vs $N$, for $U=2.50$ 
and $3.00$, the lines represent 
$I_{1}\sim N^{-1-\delta _{1}}\exp (-N/\xi _{1})$, with 
$\xi _{1}$ and $\delta _{1}$ dependent on $U$.

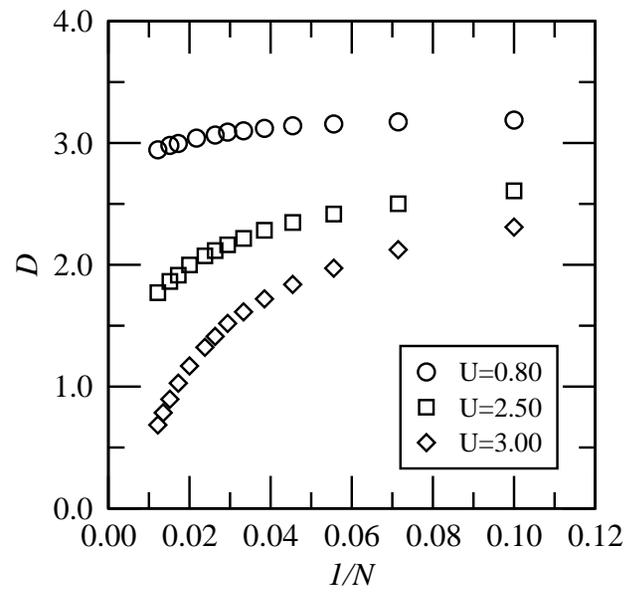
\noindent FIG. 9. The Drude weigth $D$ as a function of $1/N$, 
for different $U$ and $\rho =0.50$.

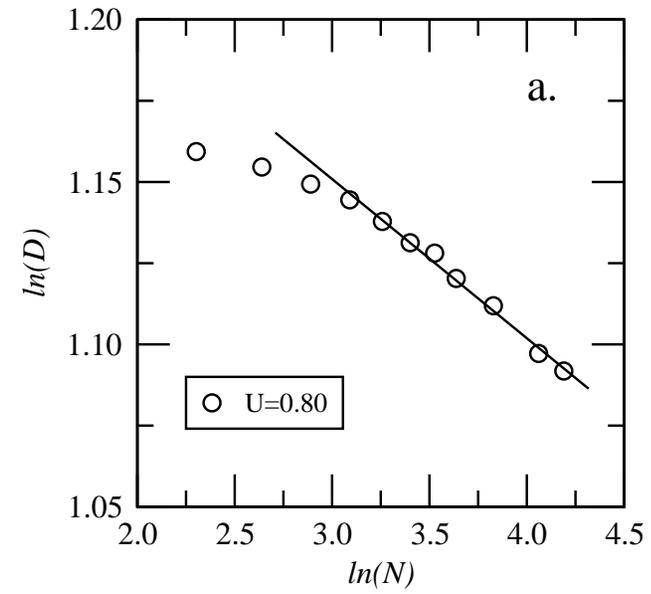
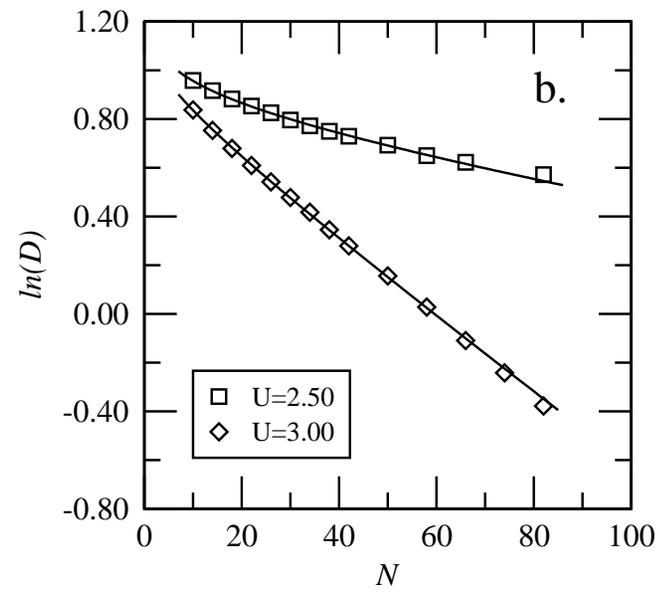
\noindent FIG. 10. $D$ for different $U$ and $\rho =0.50$. (a) 
$\ln (D)$ vs $\ln (N)$, for $U=0.80$, the line represents 
$D\sim N^{-\alpha }$; 
(b) $\ln(D)$ vs $N$, for $U=2.50$ and $3.00$, the lines represent 
$D\sim N^{-\delta }\exp (-N/\xi )$, with $\xi $ and $\delta $ dependent 
on $U$.

\end{document}